\documentclass[a4paper]{article}
 
\usepackage{bm} 
\usepackage{amssymb,amsfonts}  
\usepackage{abstract}
\usepackage{color}
\usepackage[pdftex]{hyperref}

\begin{document}

\title{{\bf{\huge{Constructive Foundation of Quantum Mechanics}}} 
}

\author{\bf{Walter Smilga}\footnote{Isardamm 135 d, D-82538 Geretsried, Germany; e-mail: wsmilga(at)compuserve.com}}

\maketitle 

\begin{abstract}
I describe a constructive foundation for Quantum Mechanics, based on the discreteness of the degrees of freedom 
of quantum objects and on the Principle of Relativity.
Taking Einstein's historical construction of Special Relativity as a model, the construction is carried out in close 
contact with a simple quantum mechanical Gedanken experiment.
This leads to the standard axioms of Quantum Mechanics.
The quantum mechanical description is identified as a tool that allows describing objects with discrete 
degrees of freedom in space--time covariant with respect to coordinate transformations.
An inherent property of this description is a quantum mechanical interaction mechanism.
The construction gives detailed answers to controversial questions, such as the measurement 
problem, the informational content of the wave function, and the completeness of Quantum Mechanics. 
\end{abstract}

\noindent
\bfseries{PACS}\mdseries: 03.65.Ta, 05.65.+b, 12.20.-m, 14.70.Bh,  \\
\bfseries{Keywords}\mdseries: Quantum foundations, interaction, self-organisation\\

\section{Introduction} 
\label{intro}  

Quantum mechanics (QM) was formulated in the years 1925 and 1926 by Born, Heisenberg, and Jordan \cite{wh,bj,bhj} 
as matrix mechanics and in the same years by Schr\"odinger \cite{es} as wave mechanics. 
QM was given its final axiomatic foundation by Dirac \cite{pd} in 1930 and von Neumann \cite{jvn} in 1932. 
Presently, in 2016, 90 years after its first formulation, there is still an ongoing discussion about the meaning 
of QM, especially about the interpretation of the wave function, which was introduced by Schr\"odinger in 1926.  
On the one hand, QM has proven to be a highly effective theory with a surprisingly large range of applications.
On the other hand, it is not inappropriate to say that its basics are largely not understood. 

In a paper of 1996, Rovelli \cite{cr} compared the present discussion about the wave function with the debates about the 
physics of length contraction and time dilation at the beginning of the 20th century.
Einstein's paper of 1905 \cite{as} suddenly clarified the situation by explaining the result of the Michelson--Morley 
experiment \cite{mm} through the Principle of Relativity \cite{hp}. 
By a simple and comprehensible Gedanken experiment (the one with the moving train) Einstein constructed a 
transformation law that allows transforming the physics in a moving frame of reference into the physics in a frame 
at rest, in accordance with the Principle of Relativity, which in this context means the constancy of the speed of light.
This transformation law is now well-known under the name of the Lorentz transformations.  
In 1905, the Lorentz transformation was already known as a coordinate transformation that leaves Maxwell's equations 
invariant. 
However, only Einstein's explicit construction unveiled its general validity and led to the formulation of the theory of  
Special Relativity (SR).
Rovelli suggested that there might be a similar constructive foundation of QM that could stop the debates about the 
correct interpretation of QM.

Following Rovelli's suggestion, I start from a simple quantum mechanical experiment, which in a way plays a similar 
role for QM as the Michelson--Morley experiment for the foundation of SR.
The experiment reveals a transformation law that, in analogy to the Lorentz transformation, allows transforming 
the results of a quantum mechanical measurement from one frame of reference to another, in accordance with the 
Principle of Relativity.
With this transformation law, the description of the experiment is not bound to a particular coordinate system: 
in other words, it allows of a covariant description.
The mathematical formulation of the transformation law leads seamlessly to the Hilbert space formalism as laid down in the 
axioms of QM. 

The explicit construction of this formalism gives detailed answers to controversial questions, such as:
What kind of information is encoded in the wave function? 
What is the role of the observer?
Is there a measurement problem?
Is the quantum mechanical description complete?
Can QM be reformulated as a deterministic theory?

\section{A Gedanken experiment}
\label{sec:1}

In the effort to understand the foundations of QM, it is reasonable to start from the most simple quantum mechanical object,
which is obviously a single spin in 3-dimensional space. 
It has only two states and neither a mass nor a charge.
The proposed experiment is a Gedanken experiment of the Stern--Gerlach type \cite{sg}, designed to measure the spin of 
an electron.
Spin is a directed quantity, which means its value is defined relative to the orientation of a given coordinate system.
Such a coordinate system can practically be realized by a measuring device capable of measuring the direction of the spin.
The result of such an experiment is well known: the spin will point in a certain direction relative to the coordinate system 
of the measuring device, or in the opposite direction, forming an angle of $0$ or $\pi$ with a given direction.
Now rotate the measuring device by an angle of $\phi$.
By this rotation, the coordinate system attached to the device is rotated by the angle of $\phi$ as well.
On the basis of classical mechanics one would expect now to observe the angles $-\phi$ and $\pi-\phi$.
Instead, the measuring device will again measure $0$ or $\pi$.
This result is in contradiction to Newtonian mechanics, but in full compliance with the Principle of Relativity,
which demands that physics be the same in every inertial system.

The Gedanken experiment can be modified by adding a second measuring device B of the same kind and considering three
steps.\\
Step 1: Measure the spin with the first device A and select all spins pointing in the direction $0$.\\
Step 2: Rotate device B by an angle $\phi$ relative to device A.\\
Step 3: Measure the spins that have passed step 1 with device B.\\
We have to examine three cases.\\
Case 1: $\phi=0$. All spins measured by B will point in the direction $0$.
This result is deterministic, insofar as from the result of the first measurement the result of the second one can
uniquely be predicted.\\
Case 2: $\phi=\pi$. All spins measured by B will point in the direction $\pi$.
Again, the result is deterministic.\\
Case 3: $0<\phi<\pi$. Now the result is non-deterministic. Device B will measure unpredictably either $0$ or $\pi$. 
However, if the experiment is repeated a sufficient number of times, well-defined probabilities for the results $0$ and $\pi$ 
can be derived. 
These probabilities depend only on the angle $\phi$ and on the spin state selected in step 1.

From the results of case (3) one can immediately find an empirical transformation law that determines how the 
results obtained in the coordinate system of device A are transformed into the coordinate system of device B.
This transformation law is non-deterministic, because it applies to probabilities rather than to states.

\section{The theory of this Gedanken experiment}
\label{sec:2}

The coordinate transformations under consideration here are continuous and invertible mappings of space--time coordinates 
onto themselves.
Cases 2 and 3 of the Gedanken experiment show that an active rotation of device B induces a non-trivial mapping of the 
measured spin states.
There is a correlation between the coordinate transformation of a (classical) measuring device and a corresponding 
transformation of the spin states.
However, whereas coordinate transformations are continuous mappings, the mapping of spin states is necessarily discontinuous.

The Principle of Relativity requires that the available spin states are the same in every coordinate system.
Hence, every invertible mapping of discrete spin states is a mapping onto themselves, and, because of the finite number of 
states, it is a permutation.
This precludes any deterministic, i.e.\,\,one-to-one relation between (continuous) coordinate transformations and 
(discontinuous) mappings of a spin.

This simple analysis of the Gedanken experiment shows that indeterminism is a well-founded phenomenon, when we observe objects 
with a discrete degree of freedom in space--time from the position of a classical observer. 
It is comprehensively understandable as a direct consequence of the discrete degree of freedom in combination with the 
Principle of Relativity.

Having clarified the reason for the statistical nature of the measurements, casting the verbal description of the Gedanken 
experiment into a concise mathematical form is now a purely technical (mathematical) issue.
We need a mathematical tool for describing:\\
(a) The initial value: the orientation of the spin measured in the reference frame of device A.\\
(b) The operation: the rotation of device B relative to device A. Alternatively expressed: the rotation of reference frame B 
relative to reference frame A.\\
(c) The final result: a statistical distribution determined from the measurements by device B.
The distribution depends only on the initial value and the rotation of reference frame B relative to A.
The final result has to be expressed as the probability of measuring a specific orientation of the spin.

At first, a description is needed for the rotation of measuring device B relative to the results obtained by device A, 
or, equivalently, for the rotation of the results obtained in reference frame A relative to reference frame B.
Since the orientation of a spin is a directed quantity, it is obvious to encode the result of device A by an abstract two-component 
vector that takes on one of the two discrete values 
\begin{equation}
\left( \begin{array}{c} 1 \\ 0 \end{array} \right) \;\;\;\mbox{or}\;\;\; \left( \begin{array}{c} 0 \\ 1 \end{array} \right) . \label{1}
\end{equation}
The vectors (\ref{1}) can then be used as basis vectors to generate a 2-dimensional (complex) vector space.
By defining a scalar product, the vector space becomes a {\it Hilbert space}.
On this Hilbert space, the well-known Hermitian Pauli matrices 
\begin{equation}
\sigma_1 = \left( \begin{array}{cc} 0 & \;\;   1 \\ 1 & \;\;  0 \end{array} \right),\;\; 
\sigma_2 = \left( \begin{array}{cc} 0 &       -i \\ i & \;\;  0 \end{array} \right),\;\; 
\sigma_3 = \left( \begin{array}{cc} 1 & \;\;   0 \\ 0 &      -1 \end{array} \right)    \label{2}
\end{equation}
generate transformations of the rotation group SO(3). 
In mathematical term: the Pauli matrices define a {\it representation} of the rotation group SO(3) on a 2-dimensional complex 
vector space.
The Pauli matrices can therefore be used to rotate the vector, representing the results obtained in reference frame A.
There is a one-to-one relation between rotations of the coordinate system B and rotations in the Hilbert space.
The Hilbert space is, therefore, a suitable bookkeeping device for uniquely encoding the rotation of reference frame B.

A vector of this Hilbert space---usually called a {\it state vector}, or, in the context of the Schr\"odinger equation, 
a {\it wave function}---allows a complete and concise description of the experimental setup: it encodes (a) the initial value 
and (b) the rotation of this initial state along with the rotation of device B relative to device A. 
Altogether the vector encodes the {\it initial conditions} for the subsequent measurement by device B.

In the language of QM, Hermitian matrices represent {\it observables}.
In the representation (\ref{2}) of the Pauli matrices, the basis vectors are eigenvectors of $\sigma_3$ with eigenvalues $1$ and $-1$.

It is easily verified that in Cases 1 and 2, the transformations in the Hilbert space correctly map the spin states in accordance 
with the Gedanken experiment. 
As regards Case 3, the Gedanken experiment as well as the mathematical considerations above clearly indicate that the outcome 
of the experiment is a probability.
Therefore, another mathematical tool is needed that, from the state vector, derives a number with the property of a probability. 
A probability is a number between $0$ and $1$, and the sum over all probabilities equals $1$.
Furthermore, the number depends on the state vector, but is a scalar under rotations of the whole of the Hilbert space.
Gleason \cite{ag} has shown that such a number is uniquely determined, once the Hilbert space is given (Gleason's Theorem 1957).
The rule that determines this number has become known as the Born rule \cite{mb}.
Born's rule states that given a state vector $\left|\psi\right\rangle$, the probability of measuring a certain eigenvalue $n$ is 
given by the squared modulus of the scalar product $|\left\langle\phi_n\mid\psi\right\rangle|^2$, where $\left|\phi_n\right\rangle$ 
is the eigenvector corresponding to the eigenvalue $n$.

Gleason's theorem defines a probability that interpolates between the pure states. 
This probability depends only on the orientation of the state vector relative to a basis vector. 
The fact that Born's rule has been experimentally verified can be considered as a clear indication that the probabilities are not
influenced by hypothetical ``hidden parameters''.
 
The abstract rotation in the Hilbert space together with Born's rule describes a transformation of the information obtained 
in reference frame A into the corresponding information that is or will be obtained in reference frame B.
This is the {\it transformation law} promised in the Introduction.

The derived formalism is identical to the usual quantum mechanical description of a spin.
Accordingly, the quantum mechanics of a spin can be understood as a descriptive tool that is independent of a specific coordinate 
system: in other words, it is a covariant description of a spin.
Nothing has entered into the setup of this description other than the empirical fact of the discreteness of spin orientations and the 
constraints imposed by the Principle of Relativity.

The same construction can be applied to a classical rigid body in place of the spin.
Since a classical body can take every orientation in space, there is a one-to-one (i.e.\,\,deterministic) relation between the 
rotation of the frame of reference and the corresponding orientation of the rigid body.
Therefore, the construction reproduces the usual description of a body by classical coordinates.
Being created by the same construction, a state vector can be understood as a generalized coordinate adapted to the needs of a 
quantum object.

Note that the coordinates of a quantum object are by construction not attributes of the quantum object itself, but rather of the
combined system of object and observer.
It is often forgotten that the same applies to classical coordinates.
The position of a classical body in ``absolute space'' is as meaningless as the idea that the ``wave function of an electron''
is an intrinsic part of the electron.

\section{Spin in space--time}
\label{sec:3}

The description of a spin developed in the previous section clearly has the well-known properties of a quantum mechanical 
description.
However, the description is not yet completed.
Since the behaviour of the spin under rotations in space identifies the spin as an object in space (and time), the
Principle of Relativity demands that the description be covariant not only with respect to rotations but also with respect 
to boost operations and translations.
Boost operations interact with rotations and vice versa.
Therefore, they cannot be treated independently.
Hence, the encoding of the boost operations must be based on the same Hilbert space formalism as already employed in 
encoding rotations.  
This means extending the spin representation of SO(3) to a spin representation of the inhomogeneous Lorentz group (Poincar\'e 
group) P(3,1) along with a doubling of the spin components to form a 4-component Dirac spinor.
The doubling of the spin components reflects the two different ways of embedding a 3-dimensional spin into 4-dimensional 
space--time, phenomenologically described as {\it particle} and {\it anti-particle}.

This extension makes use of the operations of the Poincar\'e group on the Hilbert space of a spin representation, as described in
many text-books (cf., e.g., \cite{sssch}). 
It is straightforward.
By the action of a boost operation $U(\vec{p})$ with 3-momentum $\vec{p}$ on a given spin eigenstate 
in the rest frame $\left|\psi(s)\right\rangle$ 
\begin{equation}
U(\vec{p}) \left|\psi(s)\right\rangle =: \left|\psi_s(\vec{p})\right\rangle , \label{3}
\end{equation}
a state $\left|\psi_s(\vec{p})\right\rangle$ with momentum $\vec{p}$ is generated.
(This state is no longer a spin eigenstate.)
The operator $P_\mu$ of 4-momentum is defined by
\begin{equation}
P_\mu \left|\psi_s(\vec{p})\right\rangle = p_\mu \left|\psi_s(\vec{p})\right\rangle ,\;\;\;\mu = 0,1,2,3,  \label{31}
\end{equation}
where $p_0$ can be determined from the relation $p^\mu p_\mu = m^2$, once the mass $m$ is given.
 
Any irreducible representation of P(3,1) is characterized by a mass and the modulus of an angular momentum.
In a spin representation, the states are the solutions of the Dirac equation
\begin{equation}
(\gamma^\mu p_\mu + m) \psi_s(p) = 0 \; .   \label{9}
\end{equation}
The eigenvalues $+1$ and $-1$ of the spin matrix 
\begin{equation}
\gamma^0 = \left( \begin{array}{cccc} 1 & \;\; 0 & \;\;  0   & \;\;  0 \\ 
                                      0 & \;\; 1 & \;\;  0   & \;\;  0 \\ 
																			0 & \;\; 0 &      -1   & \;\;  0 \\ 
																			0 & \;\; 0 & \;\;  0   &      -1 \end{array} \right)
\end{equation}
are responsible for the gap of $2m$ between positive and negative energies.
The eigenvalues, therefore, define a natural (dimensionless) scale for energy and mass, analogous to the scale of spin 
quantum numbers---a basic prerequisite for every discrete mass spectrum.
(When measured relative to the international prototype kilogram, a mass is assigned a dimensioned value.
If the particle, described by the Dirac equation (\ref{9}), is identified with the electron, this value is 
$9.10938356 · 10^{-31}$~kg \cite{cod}.)

By exponentiation, the operators $P_\mu$ generate translations of the momentum eigenstates $\left|\psi_s(\vec{p})\right\rangle$
\begin{equation}
U(x) \left|\psi_s(\vec{p})\right\rangle = e^{ix^\mu P_\mu} \left|\psi_s(\vec{p})\right\rangle 
= e^{i x^\mu p_\mu} \left|\psi_s(\vec{p})\right\rangle.  \label{4}
\end{equation}
This relation allows writing $P_\mu$ as a differential operator ($\hbar = 1$ throughout this paper),
\begin{equation}
P_\mu = -i \partial / \partial x^\mu , \;\;\; \mu = 0,1,2,3.           \label{5}
\end{equation}
By superposition of momentum eigenstates with appropriate phases, states can be formed that are localized at a position 
$\vec{x}$ in 3-dimensional space:
\begin{equation}
\left| \psi_s(\vec{x}) \right\rangle = \int d^3p\;e^{ip_i x^i} \left| \psi_s(\vec{p}) \right\rangle .  \label{6}
\end{equation}
(The integral is not Lorentz covariant; all that matters is that this integral can be formed in every inertial system.)
The corresponding position operators $X_i$ are defined by
\begin{equation}
X_i \left|\psi_s(\vec{x})\right\rangle = x_i \left|\psi_s(\vec{x})\right\rangle , \;\;\; i = 1,2,3.  \label{7}
\end{equation}
From Equations (\ref{5}) and (\ref{7}) the commutation relations between the position and momentum operators
\begin{equation}
\left[X_i, P_k\right] = i\,\delta_{ik}       \label{8}
\end{equation}
can be obtained. 

This finishes the provision of the tools that are required for a covariant description of a spin in momentum space and in 
configuration space.
Somewhat unexpectedly, these tools give a spin the properties of a massive spin-{\small{1/2}} particle in space--time.
Again, these properties cannot be attributed to the spin itself: they are properties as perceived by an observer in space--time.
It should be added that O'Hara \cite{ph} has presented a proof, based on Bell's inequality, that the states 
$\left|\psi_s(\vec{p})\right\rangle$ satisfy Fermi--Dirac statistics.

The provided tools are identical to those that are defined by the standard axioms of QM \cite{axioms}:\\
Axiom 1: The properties of a quantum system are completely defined by specification of its state vector $\left|\psi\right\rangle$. 
The state vector is an element of a complex Hilbert space $H$, called the space of states.\\
Axiom 2: With every physical property A (energy, position, momentum, angular momentum, ...) there exists an associated
linear, Hermitian (self-adjoint) operator $A$ (usually called an observable), which acts on the space of states. 
The eigenvalues of this operator are the possible values of its corresponding physical property.\\
Axiom 3: If $\left|\psi\right\rangle$ is the vector representing the state of a system and if $\left|\phi\right\rangle$ represents another 
physical state, there exists a probability $p(\psi, \phi)$ of finding $\left|\psi\right\rangle$ in state $\left|\phi\right\rangle$, 
which is given by the squared modulus of the scalar product on $H$: $p(\psi, \phi) = |\left\langle\phi\mid\psi\right\rangle|^2$.

A fourth axiom of QM states that for a closed system the state vector at time $t$ is derived from the state vector at time $t_0$ by a unitary 
evolution operator $U(t,t_0)$: $\left|\psi(t)\right\rangle = U(t,t_0) \left|\psi(t_0)\right\rangle = e^{i t P_0} \left|\psi(t_0)\right\rangle$.
This axiom follows immediately from relation (\ref{4}).

\section{Product states, entanglement and interaction}
\label{sec:4}

A fifth axiom of QM, which I have not yet addressed, posits that the state of two independent quantum mechanical 
systems is a product state, formed from the states of the individual systems.
This axiom can be understood as merely a consequence of Axioms 1--3: it is therefore more a rule than an axiom.
Despite its apparent simplicity, this rule has consequences that are not immediately evident:
In \cite{ws1}, I have studied some properties of irreducible two-particle representations of the Poincar\'e group.
I have shown that in these representations the individual particle states are momentum entangled, and therefore exhibit an 
interaction mediated by the exchange of momentum.
The coupling constant of this interaction can be calculated; it identifies the interaction as the electromagnetic interaction.
Hence, within a closed system of two spin-{\small{1/2}} particles, the particles behave as if they carry an {\it electric charge}.
Mathematically, this charge is not a property of the individual particles but of the specific structure of two-particle states.

The emergence of the electromagnetic interaction is remarkable, because it is derived solely from the structure of the Poincar\'e group 
under the natural constraints that in a closed two-particle system, the total momentum and angular momentum are well defined and conserved.
There is nothing here that could be called a law of nature---the electromagnetic interaction is merely a mathematical consequence of the 
quantum mechanical description.

\section{Macroscopic matter in space--time} 	
\label{sec:5}
	
Attractive and repulsive forces, as determined by the electromagnetic interaction, enable the formation of macroscopic matter, 
which in turn allows, in principle, the construction of ``classical'' measuring devices. 
In setting up the Gedanken experiment I have implicitly assumed that such devices exist.
Now this assumption is justified by unveiling the theoretical basis for the construction of these measuring devices.
This underlines the self-contained character of the constructive foundation of QM.

The electromagnetic interaction allows mapping the coordinate spaces of individual particles onto each other in an 
experimentally verifiable way.
This mapping creates a common configuration space, which we perceive as space--time.
In space--time, electromagnetic forces enable matter to form individual and distinguishable composite structures 
(and, finally, ourselves). 
Such structures can be observed, again using the electromagnetic interaction, by our visual and tactile senses.
This explains the outstanding significance of configuration space, in contrast to momentum space, for our perception 
and description of Nature.

The constructive foundation of QM treats a state in Hilbert space as a generalized coordinate of a quantum object in 
a classical background space--time continuum.
Such ``quantized'' coordinates form quantized space--time structures, similar to the quantization of space--time in various 
failed efforts to set up a theory of quantum gravity.
Note, however, that in the present context, quantized structures do not result from a ``quantization'' of the 
background continuum, but rather from the quantum properties of matter embedded in the background continuum.
Similar to classical coordinates, the coordinates of quantum objects are affected by geometric boundary conditions, 
e.g., by pinholes, double slits, apertures in scattering experiments, or, generally, by the geometry of mass distributions.
In the classical limit, due to the inherent interaction property of two-particle states, these quantized coordinates 
should lead naturally to trajectories that are not straight but curved lines in the background continuum.
This may open up a path to quantum gravity that does not require the problematic quantization of space-time.

\section{Controversial questions of Quantum Mechanics}
\label{sec:6}

From the perspective of the Gedanken experiment, the questions posed in the Introduction can be answered as follows: 

What kind of information is encoded in the wave function?\\
Firstly, the initial value of the quantum mechanical object, as obtained from a first measurement with a device A;
secondly, an operation, e.g.\,the rotation of a second measuring device B relative to device A.
From this information, the probability of a specific result is determined by applying the Born rule.

What is an observer?\\
An observer is a (classical) measuring device used to measure the state of a quantum object.
A measuring device is defined operationally as a ``device that measures'', e.g., the direction of a spin.
Compare this with SR, where a clock is defined as a ``device that measures time''.
It is certainly not practical to incorporate the internal mechanism of a certain clock or of a spin-measuring device 
into a formalism that is intended for general use.

Is there a measurement problem?\\
The wave function encodes the initial conditions of an experiment, but neither the process of measurement, nor 
the result of the measurement.
The probability of a specific result is obtained by applying Born's rule to the wave function.
Although the result of the measurement may be used to set up a new wave function, there is neither an indication 
nor a need for a physical process that lets the wave function ``collapse'' to the new wave function.

Is the quantum mechanical description complete?\\
QM is complete in the sense that the state vector (wave function) encodes all relevant information about the initial 
conditions of an experiment.
There are no ``hidden parameters'' that might might influence the experimental results, because the probability of a 
specific result is determined completely by the information contained in the state vector.

Can QM be reformulated as a deterministic theory?\\
The non-deterministic character of QM results directly from the discreteness of the states of quantum objects and the 
Principle of Relativity. 
Therefore, any deterministic reformulation of QM is likely to violate the Principle of Relativity.
The indeterminism of QM is not an indication of missing information, but merely a consequence of the discrete 
number of degrees of freedom of quantum objects.

\section{Concluding remarks} 
\label{sec:7}

The objective of this paper was to demystify the mathematical tools that govern QM.
I have constructed the tools of QM along with a simple and transparent Gedanken experiment.
The construction is based on two pillars: (1) The Principle of Relativity, and (2) the very property of a 
quantum object that gives it its name, `quantum': the discreteness of its degrees of freedom.
As a welcome by-product of this construction, the cause of the non-deterministic nature of QM has been clarified.

The construction leads to a description of quantum objects by state vectors, which can be understood as generalized 
coordinates.
Such a description is, in principle, not different from the description of a classical body by classical coordinates
in space--time. 
The quantum mechanical coordinates are adapted to the reduced number of degrees of freedom of the objects that 
they are intended to describe.
This makes them mathematically different, but by no means mysterious.

Both types of coordinates are initially no more than simple mathematical tools.
It is therefore remarkable that from the quantum mechanical description a realistic theory of interaction can be derived without 
additional laws or principles, simply by applying the quantum mechanical description to a two-particle configuration. 
This example suggests that at the ``core of physics'' we may not find any physical law, but only the seeds, 
possibly in the form of spins, for a process of self-organization that lets physical laws and physical constants emerge as 
properties of specific configurations of the seeds.
Some similarities to Wheeler's visionary ideas, as epitomized by the phrase ``It from bit'' \cite{jaw}, are undeniable.


\begin{thebibliography}{99}


\bibitem{wh}
Heisenberg W.: \"Uber quantentheoretische Umdeutung kinematischer und mechanischer Beziehungen, Zeitschrift f\"ur Physik 33 879 (1925)

\bibitem{bj}
Born M., Jordan P.: Zur Quantenmechanik, Zeitschrift f\"ur Physik 34 858  (1925)

\bibitem{bhj}
Born M., Heisenberg W., Jordan P.: Zur Quantenmechanik II, Zeitschrift f\"ur Physik 35 8  (1926)

\bibitem{es}
Schr\"odinger E.: Quantisierung als Eigenwertproblem, Annalen der Physik 79 361, 489, 734 (1926), and 81 109  (1926)

\bibitem{pd}
Dirac P. A. M.: The Principles of Quantum Mechanics, Oxford University Press, Oxford (1930)

\bibitem{jvn}
von~Neumann J.: Mathematische Grundlagen der Quanten Mechanik, Springer, Berlin (1932)

\bibitem{cr} 
Rovelli C.: Relational Quantum Mechanics, Int. J. of Theor. Phys. 35 1637 (1996)

\bibitem{as}
Einstein A.: Zur Elektrodynamik bewegter K\"orper, Annalen der Physik 17 891 (1905)

\bibitem{mm}
Michelson A. A., Morley E. W.: On the Relative Motion of the Earth and the Luminiferous Ether, American Journal of Science 34 333 (1887)

\bibitem{hp}
Poincar\'e H.: L'\'etat actuel et l'avenir de la physique math\'ematique, Bulletin des Sciences Math\'ematiques 28 302 (1904)

\bibitem{sg}
Gerlach W., Stern O.: Der experimentelle Nachweis der Richtungsquantelung im Magnetfeld, Zeitschrift f\"ur Physik 9 349 (1922)

\bibitem{ag}
Gleason A. M.: Measures on the Closed Subspaces of a Hilbert Space, Journal of Mathematics and Mechanics 6 885 (1957)

\bibitem{mb} 
Born M.: Zur Quantenmechanik der Sto{\ss}vorg\"ange, Zeitschrift f\"ur Physik 37 863 (1926)

\bibitem{sssch} Schweber S. S.: 
An Introduction to Relativistic Quantum Field Theory, Harper \& Row, New York (1962)  

\bibitem{cod} 2014 CODATA recommended values

\bibitem{ph} 
O'Hara P.: Bell's Inequality, the Pauli Exclusion Principle and Baryonic Structure. In:
Spin 96 Proceedings, World Scientific, Singapore (1997) \\
\href{http://arxiv.org/pdf/hep-th/9701089v1.pdf}{arXiv:hep-th/9701089}

\bibitem{axioms}
Lecture notes, 
\href{http://ocw.mit.edu/courses/nuclear-engineering/22-51-quantum-theory-of-radiation-interactions-fall-2012/lecture-notes/MIT22_51F12_Ch3.pdf}
{ocw.mit.edu/courses}

\bibitem{ws1} 
Smilga W.: Momentum entanglement in relativistic quantum mechnanics,\\
J. Phys.: Conf. Ser. 597 012069 (2015) \\
\href{http://iopscience.iop.org/article/10.1088/1742-6596/597/1/012069/pdf}{iopscience.iop.org/1742-6596/597/1/012069} 



\bibitem{jaw} 
Wheeler J. A.: Information, Physics, Quantum: The Search for Links. In: 
Zurek W., ed., Complexity, Entropy, and the Physics of Information, Addison-Wesley, Reading, MA (1990)

\end{thebibliography}
\end{document}